\documentclass{jpp}
\usepackage{graphicx}
\usepackage{epstopdf, epsfig}
\usepackage[]{units}
\usepackage{amsmath}

\shorttitle{Trapped-Electron Runaway Effect}
\shortauthor{E. Nilsson, J. Decker, N.~J.~Fisch and Y. Peysson}

\title{Trapped-Electron Runaway Effect}

\author{E. Nilsson\aff{1}, J. Decker\aff{2}, N.~J.~Fisch\aff{3} and Y. Peysson\aff{1}}
\affiliation{\aff{1}CEA, IRFM, F-13108, Saint-Paul-lez-Durance, France
\aff{2}Ecole Polytechnique F\' ed\'erale de Lausanne (EPFL), Centre de Recherches en Physique des Plasmas (CRPP), CH-1015 Lausanne, Switzerland
\aff{3}Princeton Plasma Physics Laboratory, Princeton University, Princeton, New Jersey 08543, USA}

\begin{document}

\maketitle

\begin{abstract}
In a tokamak, trapped electrons subject to a strong electric field
cannot run away immediately, because their parallel velocity does
not increase over a bounce period. However, they do pinch towards
the tokamak center. As they pinch towards the center, the trapping
cone becomes more narrow, so eventually they can be detrapped and
run away. When they run away, trapped electrons will have very a different
signature from circulating electrons subject to the Dreicer mechanism.
The characteristics of what are called \textit{trapped-electron runaways}
are identified and quantified, including their distinguishable perpendicular
velocity spectrum and radial extent. 
\end{abstract}

\section{\label{sec:level1}Introduction}

Under acceleration by a constant toroidal electric field, circulating
electrons in tokamaks run away when their velocity parallel to the
magnetic field is so large that the frictional collision forces become
too small to impede acceleration by the electric field. The runaway
velocity is the demarcation velocity: the electric field cannot prevent
electrons slower than this velocity from slowing down further due
to collisions; collisions are too weak to prevent electrons faster
than this velocity from undergoing acceleration by the electric field to
even higher energies. This demarcation velocity is called the critical
velocity \citep{dre59}. However, since collisions are random events,
it is not quite precisely put to term an electron as a runaway or
not. A more precise description would be to assign to each electron
a probability of running away, based on its initial set of coordinates
\citep{fisch_86,kar86a}. To the extent that this probability is a
sharp function of velocity, the notion of a critical velocity then
becomes useful.

Runaway electrons are sensitive, in addition to collisions, to other
energy loss mechanisms such as the synchrotron radiation reaction
force \citep{sta13}. They are also sensitive to perturbation of
the magnetic field, which lead to enhanced transport \citep{zen13}.
In addition the runaway population that would arise in a tokamak
magnetic field configuration is diminished owing to magnetic trapping
effects in a non-uniform magnetic field, since trapped electrons can
not immediately contribute to the runaway electron population \citep{nil14}.
The fate of the suprathermal electrons, whether trapped or circulating,
is determined from the balance of the accelerating electric field
and various radiative, convective and diffusive loss mechanisms.

Runaway populations can be quite deleterious to the operation of a
tokamak. In the case of a disruption, the loop voltage spikes, so
that large numbers of runaway electrons reach relativistic velocities
and damage the tokamak wall. Various means of mitigating the runaway
damage have been suggested, since concerns are increasing as tokamaks
become larger and carry more current, like in ITER \citep{hen07,izz11,Paz-Soldan}.
It is clearly important to understand the behaviour of the runaway
electrons in order to optimize a runaway electron mitigation strategy.

The runaway electron population in tokamaks typically arises from
the acceleration of electrons with large parallel velocities, on the
order of the critical velocity, and average perpendicular velocities,
on the order of the thermal velocity. These electrons born via the
Dreicer mechanism are circulating. In addition, a knock-on collision
between an existing runaway and slow electrons can result in two runaway
electrons. This secondary runaway generation process can generate
an avalanche of runaway production and dominate the Dreicer effect,
in particular during disruptions. These large angle collisions between
runaways and slow electrons can result in electrons with perpendicular
energies on the order of the parallel energies. When the knock-on
runaways are created near the magnetic axis, trapped particle effects
are not important \citep{ros97,Parks99,eri03}. However, these electrons
will be trapped if they are created far enough away from the magnetic
axis. In fact, since circulating runaway electrons tend to move substantially
radially outwards as they are accelerated by the dc electric field
\citep{gua10}, the secondary electrons generated via collision with
these runaways may not be near the magnetic axis at all and could
be trapped. Alternatively, a significant population of suprathermal
trapped electrons can be generated via interaction with electron cyclotron
waves.

The present paper focuses on electrons with a large enough parallel
velocity to run away in a tokamak, but are magnetically trapped because
of their large perpendicular velocity, and therefore incapable of
running away. These trapped electrons drift radially inwards due to
the Ware pinch \citep{war70}. Nearer the magnetic axis the trapping
cone contracts such that these electrons could be detrapped and run
away. When they run away, they will have a distinct signature, which
is identified in this work.

The paper is organized as follows: In Sec.~\ref{sec:2}, we describe
how trapped runaways are generated and provide their phase-space characteristics.
In Sec.~\ref{sec:3}, we discuss collisional effects on the trapped
runaways. In Sec.~\ref{sec:4},
we offer perspectives on runaway positrons
and runaway interaction with RF current drive.

\section{\label{sec:2}Signature of Trapped Runaways}

Let us describe more precisely the expected signature of the trapped
runaways. As opposed to circulating electrons, trapped electrons cannot
run away while they remain on the same flux surface, because their
interaction with the electric field results in no net gain in parallel
velocity over a bounce period. However, according to the conservation
of canonical angular momentum, they pinch towards the tokamak center.
As they pinch towards the center, the trapping condition changes such
that eventually they do run away. We call this the \textit{trapped-electron
runaway effect}, by which we refer to electrons that were initially
trapped, before running away. However, when they do run away, trapped
electrons will have a very different signature from the circulating
electrons that run away in several ways.

First, the initially-trapped runaway electrons will run away closer
to the magnetic axis of the tokamak than they were initially. Second,
the trapped electron runaways will have a distinct pitch angle corresponding
to the detrapping condition at the radial location where they run
away, which implies a high perpendicular velocity on the order of
the critical velocity. In addition, upon application of a dc electric
field these initially trapped electrons first need to undergo the
Ware pinch and associated detrapping before running away. This process
creates a delay in the runaway generation process, which can occur
only if not disrupted by collisions or other mechanisms in the meantime.

For an initially Maxwellian distribution function, the fraction of
trapped-electron runaways compared to Dreicer runaways would be small,
although it increases with the effective charge. However, the relative
importance of trapped-electron runaways might increase significantly
for three reasons: first, the usual runaways may not be well-confined,
whereas, the trapped-electron runaways, since born nearer the magnetic
axis, are very well confined; second, knock-on collisions between existing
runaways and thermal electrons produce a non-thermal population of
secondary runaways of which a significant number may be trapped due to their high perpendicular velocities \citep{nil14}. Third,
tokamak plasmas are typically non-Maxwellian, and a significant population
of suprathermal electrons with high perpendicular momentum can be
created, for instance, via resonant interaction with electron cyclotron
waves.

Assuming a circular plasma and neglecting the radial excursion of
electron orbits, the unperturbed motion of an electron guiding center
can be characterized by its radial position $r$, its momentum $p$,
and the value $\xi_{0}$ of its pitch angle cosine $\xi=p_{\Vert}/p$
on the outboard midplane where $\theta=0$ and the magnetic field
is at a minimum. Here $p_{\Vert}$ is the momentum projected in the
direction of the magnetic field. Electron trapping derives from the
adiabatic invariance of the magnetic moment $\mu=p^{2}(1-\xi^{2})/(2mB)$
along the particle orbit, such that electrons in a non-uniform magnetic
field $B(r,\theta)=B_{0}(r)/(1+\epsilon\cos\theta)$ are trapped if
$|\xi_{0}|<\xi_{T}(r)$ with
\[
\xi_{T}(r)=\sqrt{\frac{2\epsilon}{1+\epsilon}}
\]
where $\epsilon=r/R_{0}$ is the local inverse aspect ratio. 

Due to the conservation of toroidal canonical momentum in an axisymmetric
configuration, all trapped particles orbits subject to a toroidal
electric field $E_{\phi}$ drift towards the plasma center according
to the Ware pinch \citep{war70} 
\begin{equation}
\frac{dr}{dt}=-\frac{E_{\phi}}{B_{\theta}}, \label{eq:ware-1}
\end{equation}
where $B_{\theta}$ is the poloidal magnetic field. An otherwise unperturbed
electron initially trapped on the flux surface $r$ with $|\xi_{0}|<\xi_{T}(r)$
will drift inwards to the surface $r'$ where $\xi_{T}(r')=|\xi_{0}|$.
There, it will be detrapped and can run away if its parallel velocity
is above the critical value. The electron will thus become circulating
at the radial location $\epsilon'$ with 
\begin{equation}\label{eq:epsi}
\epsilon'=\frac{\xi_{0}^{2}}{2-\xi_{0}^{2}}.
\end{equation}

\begin{figure}
  \centerline{\includegraphics[scale=0.32]{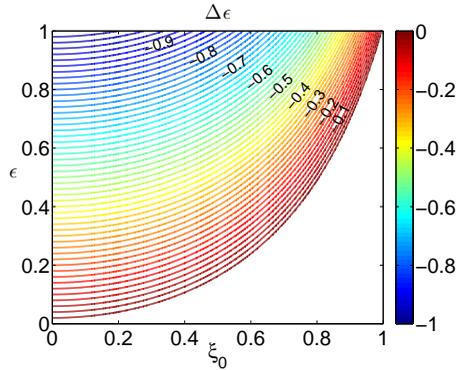}}% Images in 100% size
  \caption{The inward radial displacement ($\Delta \epsilon$) required for trapped electron initially at radial position $\epsilon$ and pitch angle $\xi_0$ to become circulating.}
 \label{fig:The-radial-displacement}
\end{figure}

\begin{figure}
  \centerline{\includegraphics[scale=0.35]{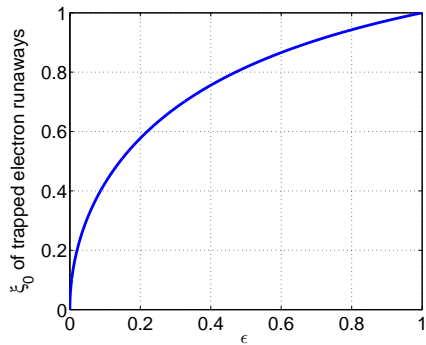}}% Images in 100% size
  \caption{The trapped-electron runaways will appear with a distinct pitch angle ($\xi_0$) in the radial direction.}
 \label{fig:xi0beam}
\end{figure}

Figure \ref{fig:The-radial-displacement} shows the required displacement
$\Delta\epsilon=\epsilon-\epsilon'$ and Fig. \ref{fig:xi0beam} the
radial position $\left(\epsilon'=r/R\right)$ where the electrons
can detrap and run away.

\begin{figure}
  \centerline{\includegraphics[scale=0.35]{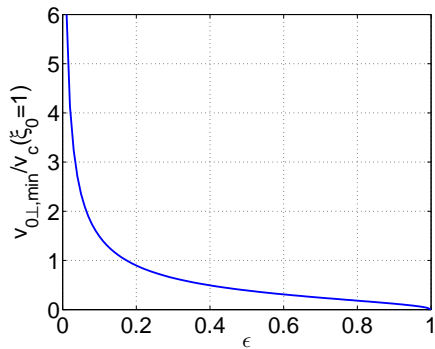}}% Images in 100% size
  \caption{The radial distribution of the minimum perpendicular velocity of the trapped electron runaways, normalised to the critical velocity for $\xi_0=1$.}
 \label{fig:vperp}
\end{figure}

Runaway electrons born at a given radial location via this detrapping
process driven by the Ware pinch will have a specific pitch-angle
according to the local trapping condition. The minimum perpendicular velocity of the trapped electron runaways is presented in Fig. \ref{fig:vperp}, normalised to $v_c({\xi_0=1})$. The time it takes for these
initially trapped electrons to become runaways is derived from the
Ware pinch velocity and radial displacement until detrapping occurs

\begin{figure}
  \centerline{\includegraphics[scale=0.3]{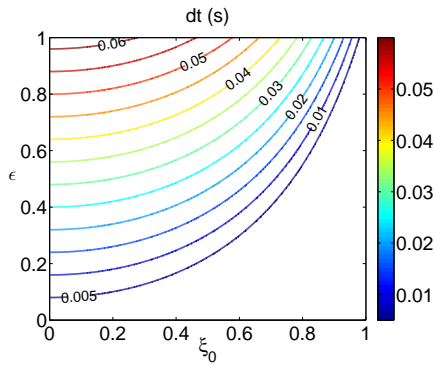}}% Images in 100% size
  \caption{Time (s) for trapped electrons at ($\xi_{0},\epsilon$) to reach the
radial position where they become passing electrons, for $B_{\theta}=\unit[0.05]{T}$, $E_{\phi}=\unit[0.8]{V/m}$ and $R=\unit[1]{m}$.}
 \label{fig:Time-(s)-for}
\end{figure}

\begin{equation}
dt_{W}=\frac{B_{\theta}}{E_{\phi}}R\left(\epsilon-\frac{\xi_{0}^{2}}{2-\xi_{0}^{2}}\right).\label{eq:t}
\end{equation}
For an equilibrium with $B_{\theta}=\unit[0.05]{T}$ and $E_{\phi}=\unit[0.8]{V/m}$.
The time required for a trapped electron to become passing is shown
in Fig.~\ref{fig:Time-(s)-for}.

\begin{figure}
  \centerline{\includegraphics[scale=0.3]{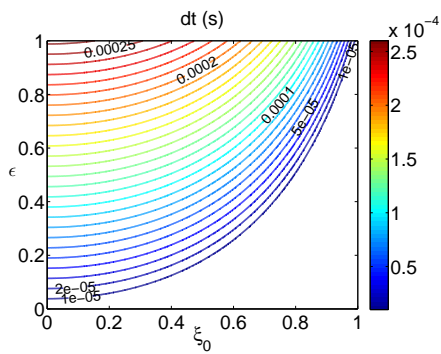}}% Images in 100% size
  \caption{Time (s) for trapped electrons at ($\xi_{0},\epsilon$) to reach the
radial position where they become passing electrons, for $B_{\theta}=0.01$
T, $E_{\phi}=38$ V/m and $R=\unit[1]{m}$.}
\label{fig:Time-(s)-for-1}
\end{figure}

In a disruption in an ITER-like scenario the toroidal electric field
can be much stronger; around $\unit[38]{V/m}$ has been predicted
\citep{hen07}. In that case, the Ware pinch detrapping time scale
would be much shorter; see Fig.~\ref{fig:Time-(s)-for-1}.

Corrections to the detrapping radius and detrapping time due to the
Shafranov shift, non-circular plasma shape, grad-B and curvature drifts
are beyond the scope of this paper. It is important to note that very
near the magnetic axis the approximations made in this paper break
down.

Although the poloidal magnetic field $B_{\theta}$ has little effect
on the detrapping condition, it does have a strong effect on the detrapping
time. For simplicity, we took the $E_{\phi}/B_{\theta}$ velocity
to be constant. In fact, as the electron pinches inward, the poloidal
field decreases, thereby speeding up the drift. Thus, a more precise
formulation of Eq. \ref{eq:t} would account for the $E_{\phi}/B_{\theta}$
dependence upon $r$.

\section{\label{sec:3}Limitations of the collisionless approach}

The collisionless approximation used in Sec. \ref{sec:2} is valid if the
pinch time is small compared to both the collisional slowing down
and detrapping times of runaway electrons. If the pinch time is longer
than the slowing-down time, the electron may slow down such that it
may not have the energy required to run away when it finally detraps.
If the electron undergoes significant pitch-angle scattering during
the pinch time, it may be detrapped at a different radial location.

Since the collision time increases with velocity, the validity condition
for the collisionless approch must be evaluated from the pitch angle
dependent critical momentum \citep{ros97} 
\begin{equation}
p_{c}^{2}\approx\frac{2}{1+\xi_{0}}\frac{E_{c}}{E},\label{eq:pc-1}
\end{equation}
where the critical field $E_{c}$ is proportional to the plasma density.
The slowing-down time for electrons with velocity $v_{c}/c=p_{c}/\gamma_{c}$
is 
\[
\tau_{c}=\frac{4\pi\epsilon_{0}^{2}m_{e}^{2}v_{c}^{3}}{q^{4}n_{e}\ln\Lambda}.
\]
 Slowing-down can thus be neglected if $dt_{W}\ll\tau_{c}$. 

Collisional detrapping via pitch-angle scattering occurs over a shorter
time than the collision time that is proportional to the square of the width $\xi_{T}$ of the trapping region and can be estimated as 
\[
\tau_{dt}\sim\frac{\epsilon\tau_{c}}{1+Z_{\textrm{eff}}}.
\]

If the condition $dt_{W}\ll\tau_{dt}$ is not satisfied, trapped-electron
runaways will still be generated, but the detrapping radial distribution
will be different. 

\begin{figure}
  \centerline{\includegraphics[scale=0.32]{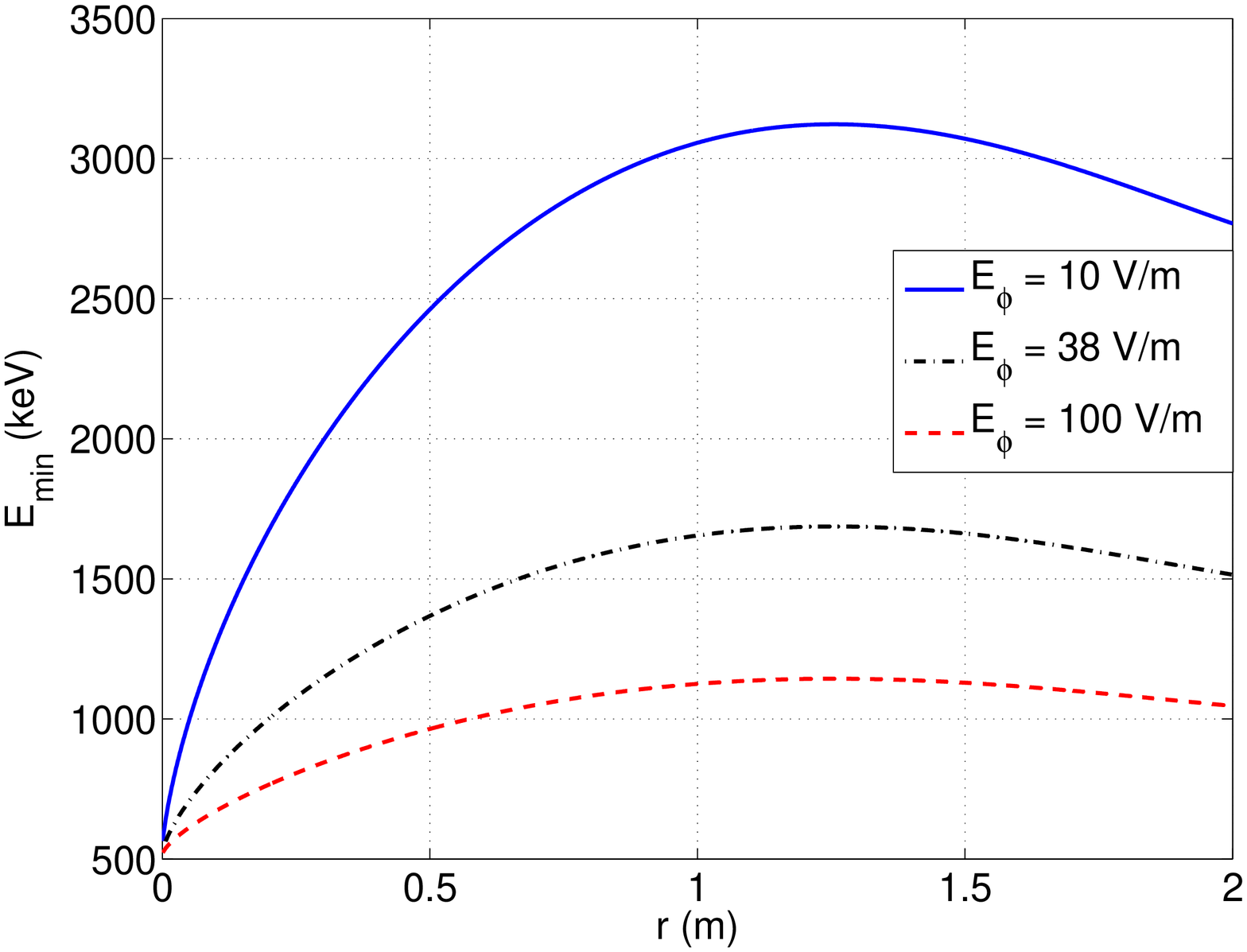}}% Images in 100% size
  \caption{The minimum energy needed for the Ware pinch to be faster than the collision time, for the parameters $E_\phi=\unit[10]{V/m}$, $\unit[38]{V/m}$ and $\unit[100]{V/m}$, $n_e=\unit[10^{20}]{m^{-3}}$ and $R=\unit[6.2]{m}$. The poloidal magnetic field is calculated from a current density profile with a total plasma current of $I_p=\unit[15]{MA}$.}
 \label{fig:Emin}
\end{figure}

The collisionless condition is shown in Fig.~\ref{fig:Emin} for
ITER-like parameters and various values of the electric field. The
density is $n_{e}=\unit[10^{20}]{m^{-3}}$ . The magnetic field is
calculated from a current density profile, peaked on-axis and with a total plasma current
of $I_{p}=\unit[15]{MA}$. With the current profile assumed in this
calculation, the maximum of the poloidal magnetic field is located
inward from the plasma edge, which explains the maximum in the energy
condition. The radial dependence of the density and the electric field
strength are not taken into account.

The minimum energy required for the collision time to be longer than
the Ware pinch time is on the order of $\unit{MeV}$ for ITER-like
parameters during a disruption. Considering
that runaway electrons have many tens of $\unit{MeV}$, it can be
expected that runaways produced from knock-on collisions should be
in the range where collisions can be neglected. For trapped electron
runaways with lower energy, collisions must be accounted for. That
regime is left for a future study.

Just like the collisional slowing down, the synchrotron reaction force \citep{pau58} limits the energy of the particle. 

The relativistic characteristic time for the radiation reaction force is:
\begin{equation}
\tau_r = \frac{6 \pi \varepsilon_0 \gamma (m_0 c)^3}{q^4B^2} \approx 5.2\frac{\gamma}{Z^4 B^2} .
\end{equation}

In addition the synchrotron reaction force limits the pitch angle $\theta=\arccos(\xi)$ \\\citep{hel02a}
\begin{equation}
F_{rad,\,\xi} = -\frac{p\xi \sqrt{1-\xi^2}}{\gamma \tau_r}. 
\end{equation}

This force could affect the trapped electron runaways in the Ware pinch process, as electrons would detrap faster, i.e. at a larger radius than predicted in Sec.~\ref{sec:2}, if $p_\perp / p_\|$ decreases.

Limitations of the collisionless theory was discussed previously in this section and a regime was identified where the Ware pinch detraps trapped runaways fast enough for the collisions to be negligible ($dt_W \ll \tau_c$). A similar condition can be set for the radiation loss time $dt_W \ll \tau_r$. We compare the time scale of collisional damping with the one of the radiation damping:
\begin{equation}
\frac{\tau_c}{\tau_r}= \frac{2 \epsilon_0}{3m_e n_e \ln \Lambda } \gamma \Big(\frac{v}{c}\Big)^3 Z^4 B^2 \approx \frac{\gamma}{n_{e,19}} \Big(\frac{v}{c}\Big)^3 Z^4 B_{[5T]}^2,
\end{equation}
%for $\ln \Lambda=16$ 
where $n_{e,19}$ is the electron density in the unit $\unit[10^{19}]{m^{-3}}$ and $B_{[5T]}$ in units of $\unit[5]{T}$. For relevant plasma parameters, the condition $\tau_c \ll \tau_r$ is fulfilled unless $\gamma$ gets very large. From the minimum energy defined in Fig. 6, where the Lorentz factor $\gamma$ is in the range of $1-6$, for higher $\gamma$ the time scale of the synchrotron reaction force may be short enough to change the pitch before the detrapping radius is reached, if the product $B^2 Z^4/n_e$ becomes large. Since this effect would speed up the detrapping process, the prediction in Sec.~\ref{sec:2} can be considered as an upper estimate of the detrapping time and lower estimate of the detrapping radius. To properly account for the combined effect of synchrotron reaction and collisional drag on the trapped-electron runaway distribution during the Ware pinch would require further investigation by numerical studies.

\section{\label{sec:4}Discussion}

This paper describes how initially trapped electrons may become runaway
electrons if their parallel velocity is above the critical runaway
velocity as they become detrapped following the inward Ware pinch.
These runaway electrons are born nearer the magnetic axis as compared
to their initial location, and with a high perpendicular velocity
corresponding to the trapped/circulating boundary. That will distinguish the trapped runaways from the passing runaways, which have average perpendicular energies. They will produce
a relatively more intense synchrotron radiation than Dreicer runaways.
The production dynamics of the trapped-electron runaways is determined
by the Ware pinch time. There will be a turn-on time for the electrons to reach the radial detrapping position and only then to begin to run away. The presence of trapped-electron runaways may affect the radial profile of runaway electrons since they are concentrated near the magnetic axis. Even in the case where primary generation would be small in the center, for example for very peaked density, one could still expect a centrally concentrated runaway electron population under some circumstances. This would be the case if the avalanche growth rate from knock-on collisions decays strongly off-axis owing to magnetic trapping effects as found for toroidal geometries \citep{nil14}. At the same time, the trapped-electron runaways are concentrated near the center, as found in the previous section. 

In other words, the radial dependence of the growth rate of runaways depends on various effects, where the avalanche effect and the trapped-electron runaway effect would weight the runaway distribution towards the center. Quantitative predictions of the radial profile of runaway electrons are left to future studies. 
All these signatures should be most prominent during a disruption, where the electric field is large, and might be used to provide information on plasma conditions. The combination of high synchrotron emission
and specific dynamics could make it possible to identify the signature
of trapped-runaway electrons during disruptions when a large number
of energetic trapped electrons is generated via knock-on collisions.

The large perpendicular energies of the trapped-electron runaways
also suggest that they may be easier to control than conventional
Dreicer runaways, as they can be deconfined through interactions with
ripple fields \citep{Laurent,Rax_runaway}, coherent wave instabilities
\citep{Fulop_Newton,Fulop_whistler} or magnetic perturbations \citep{papp_2011}.
In all of these processes, the strength of the interaction increases
with perpendicular momentum.

Recently, there has been interest in the creation of runaway positrons
in tokamaks, and the information that might be obtained from them
upon annihilation \citep{Helander_Ward,Fulop_Papp,liu14}. When large
tokamaks disrupt, large electron-positron pair production is expected
to occur. The positron runaways behave just like electron runaways,
only they run away in the opposite direction. Just as there are
circulating positron runaways, there will be trapped positron runaways.
These trapped positron runaways will pinch towards the tokamak magnetic
axis just like the trapped electron runaways. Except that they will travel
in the opposite toroidal direction, which will affect the Doppler
shift of the synchrotron radiation, the trapped positron runaways
will have a completely analogous signature to the trapped electron
runaways. Moreover, since the positrons would only be produced in
large numbers through the avalanche effect involving very high energy
runaways, there will be relatively more of the trapped positron runaways
(compared to the usual positron runaways) than there would be trapped
electron runaways (compared to the usual electron runaways). This
effect would be enhanced to the extent that the most energetic runaways
-- and those most capable of the pair production -- would be found
near the low field side of the tokamak \citep{gua10}, where the
trapping effect is also most significant.

Note that the trapped-electron runaway effect that we discuss is for runaway electrons that run away eventually in the direction in which they support the plasma current. The same is true for the runaway positrons. When the tokamak current is
maintained by a dc electric field, it is after all the thermal electrons that carry the toroidal current, and the runaway electrons carry current
in the same direction, only they are accelerated to far higher energies. This would also be true during start-up of the tokamak \citep{Mueller}, if the start-up relies on an inductive current. In such a case, there is also danger from runaway electrons, since the plasma may not be so dense as to hold back the runaways.  
Moreover, in the case of RF-assisted start-up of the current, such as through electron cyclotron heating, there might be more electrons produced at higher energies, which could then run away in the direction in which the runaways support the current.

Whether or not the runaways are in a direction to  support the plasma current is an important distinction that comes into play in non-inductive start-up of the tokamak current. In the case of non-inductive current drive, for example by RF waves  \citep{fis87}, there is the opportunity to start up the tokamak or to recharge the transformer \citep{fisch_transformer}. In such a case, the loop voltage is driven negative; in other words, as the RF-current is increased, a loop voltage is induced which opposes the RF-driven current. This dc electric field also
produces runaway electrons, only now they are so-called {\it backwards
runaways}, which are runaway electrons that carry current counter to
the toroidal current  \citep{fisch_86,kar86a}. It is also important to note that the trapped-electron runaways are
not a concern in the case of backward runaways. In this case, which
may occur during the startup or flattop phases in the presence of
strong RF current drive  \citep{fisch_transformer,fis87,Karney_PRA,Li_EAST_startup,Ding2012},
the electric field opposes the plasma current such that the pinch
is directed outward where the trapping cone widens. Hence, in the case of RF ramp-up, while circulating backwards runaways are produced,
there is no production whatsoever of backwards trapped-electron runaways (or, for the same reason, backwards trapped-positron runaways).
Thus, in vigorous RF ramp-up regimes, while the circulating backwards
runaways might be of some concern, at least the trapped-electron runaways
do not add to that concern.

A recurring question is to what extent RF current drive generates runaway electrons. This question should also be posed for the trapped-electron runaways. In the case of RF current drive, if the current drive effect relies on RF wave interactions with suprathermal electrons, there is risk of producing runaway electrons. It is particularly the case for lower hybrid current drive \citep{fis78}, where a suprathermal electron tail is formed at high parallel velocities that could supply more runaway electrons than could a Maxwellian distribution. It is also the case for electron cyclotron current drive \citep{fis80}, where heating in perpendicular velocity makes electrons collide less frequently and become more likely to run away. In these cases, the RF heating of passing electrons enhances the runaway current through the usual runaway effect. However, there is also a trapped particle runaway effect when the RF current drive affects trapped electrons. Consider first waves that provide parallel momentum to energetic trapped electrons, such as low parallel-phase-velocity waves \citep{wor71}.  Since the particles remain trapped, there is an RF-pinch effect similar to the Ware pinch effect \citep{fis81}. If the wave momentum is in the direction supportive of the total current, then as with the Ware pinch effect, the pinch will be inwards. Moreover, as with the Ware pinch, the trapped electrons experience less stringent trapping conditions when they pinch, so they can eventually run away like a trapped runaway. One important difference is that, as opposed to the Ware pinch effect where the electric field pinches the electron, without increasing its energy, in the case of the RF pinch effect, the RF waves pinch the electron, while increasing its energy. As a result the trapped runaways become detrapped sooner, and so run away at larger radii. It must be pointed out that the RF-pinch may only occur if the wave-particle resonance is present continuously through the pinching process, i.e. if the spatial distribution of the waves has sufficient radial extent. In any event, in inputting parallel momentum with waves that would be supportive of the toroidal current, whereas targeting electrons with higher parallel velocity can increase the number of runaway electrons, targeting electrons with low parallel velocity can increase the number of trapped runaways.

In contrast, in the case of perpendicular heating rather than parallel heating of trapped electrons, such as by electron cyclotron waves, there is no pinch effect bringing electrons to less stringent trapping conditions. In fact, the perpendicular heating causes the electrons to be more deeply trapped.  Hence, there is no trapped-particle runaway effect for heating by electron cyclotron waves. 

\section{\label{sec:6}Summary}
To sum up, we identified the trapped-electron runaway effect. We calculated the key parameters that distinguish these runaways, namely the large perpendicular energy, the dependency of the perpendicular energy on radius, and the turn-on time for the appearance of the runaways. We identified when these effects might be expected (in the case of positrons) and when they would be absent (in the case of RF ramp-up). Possible observables would therefore be based on signals sensitive to perpendicular energy, such as synchrotron radiation. Similarly, the degree of manipulation by waves or magnetic perturbations is also sensitive to perpendicular energy. Thus, we hope that these observations and calculations will assist in formulating methods of controlling those runaways or in utilising measurements of their behaviour for informing on other processes in the plasma. 

%%%%%%%%%%%%%%%%%%%%%%%%%%%%%%%%%%%%%
\vskip 10pt
{\bf Acknowledgements}
\vskip 5pt
This work has been carried out within the framework of the EUROfusion Consortium and has received funding from the Euratom research and training programme 2014-2018 under grant agreement No 633053. The views and opinions expressed herein do not necessarily reflect those of the European Commission. The authors appreciate the hospitality of the  Chalmers University of Technology, where, at a meeting organized by  Professor T\"unde F\"ul\"op,  these ideas were initially conceived.  The authors are grateful to Eero Hirvijoki, Isztvan Pusztai and Adam Stahl for fruitful discussions. One of us (NJF) acknowledges support, in part, from DOE  Contract No.~DE-AC02-09CH11466.

\bibliographystyle{jpp}

\end{document}